\documentclass[10pt]{iopart}
\usepackage{graphicx}% Include figure files
	\graphicspath{{figures/}}
\expandafter\let\csname equation*\endcsname\relax
  \expandafter\let\csname endequation*\endcsname\relax
\usepackage{amssymb}
\usepackage{amsmath}
\usepackage[usenames, dvipsnames]{color}
\usepackage[colorlinks,
            linkcolor=blue,
            anchorcolor=blue,
            citecolor=blue,
            urlcolor=blue
            ]{hyperref}
\usepackage{siunitx}
\usepackage{physics}
\usepackage{braket}
\usepackage{bm}% bold math
\usepackage{cite}
\begin{document}

\title[]{Topological spin-Hall edge states of flexural wave in perforated metamaterial plates}

\author{Linyun Yang, Kaiping Yu$^*$, Ying Wu, Rui Zhao and Shuaishuai Liu}

\address{Department of Astronautic Science and Mechanics, Harbin Institute of Technology, Harbin, Heilongjiang 150001, China}
\ead{yukp@hit.edu.cn}
\vspace{10pt}
\begin{indented}
\item[]March 2018
\end{indented}

\begin{abstract}
This paper investigates  the pseudo-spin based edge states for flexural waves in a honeycomb perforated phononic plate, which behaves an elastic analogue of the quantum spin Hall effect. We utilize finite element method to analyse the dispersion for flexural waves based on Mindlin's plate theory. Topological transition takes place around a double Dirac cone at $\Gamma$ point by adjusting the sizes of perforated holes. We develop an effective Hamiltonian to describe the bands around the two doubly degenerated states and analyse the topological invariants. This further leads us to observe the topologically protected edge states localized at the interface between two lattices. We demonstrate the unidirectional propagation of the edge waves along topological interface, as well as their robustness against defects and sharp bends.
\end{abstract}


\vspace{-10pt}
\submitto{\JPD}
%\ioptwocol

\section{Introduction}

The discovery of topological insulators\cite{von1986quantized,hasan2010colloquium,qi2011topological}, which can exhibit topologically protected edge states propagating in a single direction along the sample edges, has opened a new chapter in the research realm of condensed matter physics. The edge states are immune to backscattering from disorder or sharp bends because of the underlying topology of the band structures. Recently these concepts in quantum systems have been extended to the field of photonic\cite{haldane2008,lu2014,Wulonghua2015}, acoustic\cite{khanikaev2015,Yangzhaoju2015,Chenzeguo2016,Hecheng2016,Meijun2016,Xiabaizhan2017,SimonYves2017,Lujiuyang2016,fleury2016,Lujiuyang2017,Zhangzhiwang2017,NiXiang2017} and elastic\cite{susstrunk2015,Wangpai2015,nash2015,huber2016,torrent2013,mousavi2015,miniaci2017,Yusiyuan2017,chaunsali2018,foehr2017,
pal2017edge,Zhuhongfei2017,Wuying2018} systems, due to their potential practical applications in wave guiding, isolating, filtering, etc.

Topologically protected states in acoustic and mechanical systems become a research focus very recently. Breaking the time-reversal symmetry and mimicking the quantum Hall effect (QHE)\cite{khanikaev2015,Yangzhaoju2015,Chenzeguo2016,Wangpai2015,nash2015} in mechanical and acoustic systems requires external active components, such as gyroscope and circulating fluid flow, which adds complexity to the systems and remains challenging to practical realization. Another scheme to achieve topological states is establishing acoustic/mechanical analogues to the quantum spin Hall effect (QSHE)\cite{mousavi2015,Hecheng2016,Meijun2016,SimonYves2017,miniaci2017,Yusiyuan2017,chaunsali2018,foehr2017,Xiabaizhan2017}, which makes use of the two irreducible representations of  $C_{6v}$ point group symmetry to construct (pseudo) spin Hall states. Forming a double Dirac cone to increase the degrees of freedom is essential to realize the acoustic/mechanical analogue of QSHE, since Kramers doublet exists in the form of two double degenerate states, pseudo spin states can be constructed by the hybridization of two pairs of degenerate Bloch modes.

Different from acoustic waves, elastic waves in plates exhibit much more complex dispersion behaviors due to the existence of both longitudinal and shear waves, and moreover, the reflections and couplings at the stress-free boundaries. Elastic plates have become an attractive platform for the study of topological states for both academic and practical interests very recently. These investigations mainly focus on the lamb modes\cite{mousavi2015,miniaci2017,Yusiyuan2017,Zhuhongfei2017} and flexural modes\cite{torrent2013,pal2017edge,chaunsali2018,foehr2017} in plates with perforated holes\cite{mousavi2015,miniaci2017,Yusiyuan2017} or resonators\cite{torrent2013,pal2017edge,chaunsali2018,foehr2017,Zhuhongfei2017}, forming elastic analogues of QSHE\cite{mousavi2015,miniaci2017,Yusiyuan2017,chaunsali2018,foehr2017} or quantum valley Hall effect (QVHE)\cite{pal2017edge,Zhuhongfei2017}. Recently, Rajesh and co-authors\cite{chaunsali2018} investigated the topological spin Hall effects for flexural waves based on local resonant metamaterial plate with \textit{mass-spring} systems attached on one face of the plate. In their work, the distance between the resonators and the unit cell center is tuned to break the translational symmetry in order to open the double Dirac cone. In this paper, inspired by zone-folding mechanism, we report the observation of topologically protected edge states for flexural waves in plates with solely honeycomb arranged \textit{circular holes}. By simply perturbing the holes' radii, the topology transition from a topological trival band to non-trival one can be realized. Our structural design strategy is free of fabrication complexity, and provides an ideal platform for practical realization for elastic QSHE analogues.

The structure of this work is outlined as follows, in section \ref{sec:2}, we present the dispersion analysis of flexural waves based on Mindlin's plate theory using finite element method (FEM). In section \ref{sec:3}, we show the band inversion process and study the topology of the band structures. In section \ref{sec:4}, we consider a ribbon-shaped supercell composed topological trival and non-trival cells, and show the existence of a pair of topological edge states. This is followed in section \ref{sec:5} by a full wave simulation to check the topological protected pseudo-spin dependent flexural wave transport. A brief summary and discussion of this paper is provided in section \ref{sec:6}.

\section{\label{sec:2}Dispersion Analysis for Perforated Phononic Plates}

\begin{figure}[htbp]
	\centering
	\includegraphics[width=0.8\linewidth]{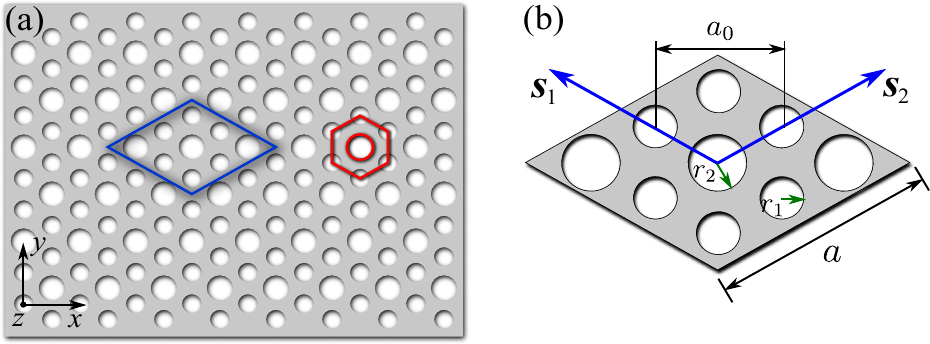}
	\caption{\textbf{Schematic view of the phononic plate and unit cells.} \textbf{(a)} Two types of unit cells  are marked by blue (rhomb) and red (hexagon) lines. \textbf{(b)} A zoom in view of the rhombic cell, $\bm{s}_1,~\bm{s}_2$ are the two lattice vectors, $a$ denotes the lattice constant, and $r_1$ and $r_2$ are the radii of the smaller- and bigger-sized perforated holes.}
	\label{fig:Fig1}
\end{figure}

We consider a flat thick plate with hexagonal arranged perforated holes, as figure \ref{fig:Fig1} shows. Two types of unit cells are highlighted by red hexagon and blue rhomb in figure \ref{fig:Fig1} (a). The hexagonal cell (primitive unit cell) can clearly illustrate the honeycomb arrangement pattern and the $C_{6v}$ symmetry of our system, but the complex boundaries may bring some inconvenience in the following analysis. Therefore, we choose a bigger rhombic shaped unit cell, shown in figure \ref{fig:Fig1} (b). $\bm{s}_1$ and $\bm{s}_2$ denote the two lattice vectors. The plate material is acrylic glass, with Young's modulus $E=3.20\si{GPa}$, Poisson's ratio $\nu=0.33$ and mass density $\rho=1062\si{kg/m^3}$. $a_0=0.05\si{m}$ and $a=\sqrt{3}a_0$ are the lattice constants of the hexagonal and rhombic unit cells, and the plate thickness is set to be $h=0.12a_0$.

Dispersion analysis for flexural waves propagating in plate structures in this paper is based on Mindlin's plate theory, which is an extension of classical Kirchhoff plate theory by taking into account the first order shear deformation\cite{achenbach_wave}. We employ FEM to calculate the dispersion relation in phononic crystal plate described above. According to the First-Order shear deformation theory for flexural vibration modes in elastic plates, the displacement can be expressed as
\begin{equation}
\left\{
\begin{aligned}
	u(x,y,z)&=z\theta_x(x,y)\\
	v(x,y,z)&=z\theta_y(x,y)\\
	w(x,y,z)&=w(x,y)
\end{aligned}
\right.
\label{eq:trunc}
\end{equation}
in which $u,v,w$ are the particle displacement components in $x,y$ and $z$ directions, and $\theta_x, \theta_y$ are the rotations of the normal to the mid-plane with respect to axis $y$ and $x$, respectively.

So the strains can be expressed as
	    \begin{subequations}
		\begin{equation}
			\varepsilon_x=z\frac{\partial\theta_x}{\partial x}, \quad
			\varepsilon_y=z\frac{\partial\theta_y}{\partial y}, \quad
			\gamma_{xy}=z\bigg(\frac{\partial\theta_x}{\partial y}+\frac{\partial\theta_y}{\partial x}\bigg)\label{eq:epsilon1}
		\end{equation}
		\begin{equation}
			\gamma_{xz}=\theta_x+\frac{\partial w}{\partial x}, \quad
			\gamma_{yz}=\theta_y+\frac{\partial w}{\partial y}
			\label{eq:epsilon}
		\end{equation}
		\label{eq:strain}
		\end{subequations}
And stresses are given by the stress-strain relation
		\begin{align}
			\sigma_f=D_f\varepsilon_f,\quad \sigma_c=D_c\varepsilon_c
		\end{align}
where $\sigma_f=[\sigma_x,\sigma_y,\tau_{xy}]^{\top}, \varepsilon_f=[\varepsilon_x,\varepsilon_y,\gamma_{xy}]^{\top}$, and $\sigma_c=[\tau_{xz},\tau_{yz}]^{\top},\varepsilon_c=[\gamma_{xz},\gamma_{yz}]^{\top}$. $D_f$ and $D_c$ are defined as

\begin{equation}
	D_f=\frac{E}{1-\nu^2}
	\begin{bmatrix}
		1 & \nu & 0\\
		\nu & 1 & 0\\
		0 & 0 & \frac{1-\nu}{2}
	\end{bmatrix},\quad
	D_c=
	\begin{bmatrix}
		\mu & 0\\
		0 & \mu
	\end{bmatrix}
\end{equation}
in which $E$ and $\mu$ are Young's modulus and shear modulus, $\nu$ is Poisson's ratio. A four-node bi-linear isoparametric element is employed to discretize the unit cell\cite{ferreira_matlab}, and displacement variables are interpolated by the nodal displacements,
\begin{equation}
	w(x,y)=\sum_{i=1}^nN_i(x,y)w_i,~\theta_x(x,y)=\sum_{i=1}^nN_i(x,y)\theta_{x,i},~\theta_y(x,y)=\sum_{i=1}^nN_i(x,y)\theta_{y,i}
	\label{eq:interp}
\end{equation}
where $n$ denote the number of nodes in each element, and $w_i,~\theta_{x,i}$ and $\theta_{y,i}$ are the displacement components at the $i$th node. $N_i(x,y)$ are shape functions that are used to interpolate the nodal displacements. Inserting equation \eqref{eq:interp} into equation \eqref{eq:strain}, we have
\begin{equation}
	\varepsilon_f = z\bm{B}_f\bm{u}^{\mathrm{e}},\quad \varepsilon_c=\bm{B}_c\bm{u}^{\mathrm{e}}
	\label{eq:strain_disp}
\end{equation}
with $\bm{u}^{\mathrm{e}}=[w_1,~\theta_{x,1},~\theta_{y,1},~,\ldots,w_4,~\theta_{x,4},~\theta_{y,4}]^{\top}$ the element nodal displacement vector. $\bm{B}_f$ and $\bm{B}_c$ are given by
\begin{subequations}
\begin{equation}
	\bm{B}_f=
	\begin{bmatrix}
		0 & N_{1,x} & 0 & \ldots & 0 & N_{4,x} & 0\\
		0 & 0 & N_{1,y} & \ldots & 0 & 0 & N_{4,y}\\
		0 & N_{1,y} & N_{1,x} & \ldots & 0 & N_{4,y} & N_{4,x}
	\end{bmatrix}
\end{equation}
\begin{equation}
	\bm{B}_c=
	\begin{bmatrix}
		N_{1,x} & N_1 & 0 & \ldots & N_{4,x} & N_4 & 0\\
		N_{1,y} & N_1 & 0 & \ldots & N_{4,y} & N_4 & 0
	\end{bmatrix}
\end{equation}
\end{subequations}
Here $N_{i,x}=\frac{\partial N_i}{\partial x}$ and $N_{i,y}=\frac{\partial N_i}{\partial y}$. The strain energy and kinetic energy in each element can be expressed as
\begin{subequations}
\begin{equation}
	U^{\mathrm{e}}=\frac{1}{2}\int_{V^{\mathrm{e}}}\varepsilon_f^{\top}D_f\varepsilon_f\mathrm{d}V+\frac{\kappa}{2}\int_V\varepsilon_c^{\top}D_c\varepsilon_c\mathrm{d}V
\end{equation}
\begin{equation}
	T^{\mathrm{e}}=\frac{1}{2}\int_{V^{\mathrm{e}}}\bm{\dot{u}}^{\top}\mathrm{diag}(\rho,I,I)\bm{\dot{u}}\mathrm{d}V
\end{equation}
\label{eq:energy}
\end{subequations}
in which $\kappa=\pi^2/12$ is the correction factor, and $\dot{u}=\frac{\partial u}{\partial t}$. Thus the Lagrangian of this system is given by $\mathcal{L}=T-U$. From the Lagrangian equation
\begin{equation}
	\frac{\mathrm{d}}{\mathrm{d}t}\frac{\partial\mathcal{L}}{\partial\dot{\bm{u}}^{\mathrm{e}}}-\frac{\partial\mathcal{L}}{\partial\bm{u}^{\mathrm{e}}}=0
\end{equation}

We can obtain that
\begin{equation}
	\mathbf{K}^{\mathrm{e}}\bm{u}^{\mathrm{e}}+\mathbf{M}^{\mathrm{e}}\ddot{\bm{u}}^{\mathrm{e}}=0
\end{equation}
where $\mathbf{K}^{\mathrm{e}},~\mathbf{M}^{\mathrm{e}}$ are so called element stiffness matrix and element mass matrix, whose expressions are given by
\begin{subequations}
\begin{equation}
	\mathbf{K}^{\mathrm{e}}=\frac{h^3}{12}\int_{A^{\mathrm{e}}}\bm{B}_f^{\top}\bm{D}_f\bm{B}_f\mathrm{d}A+\kappa h\int_{A^{\mathrm{e}}}\bm{B}_c^{\top}\bm{D}_c\bm{B}_c\mathrm{d}A
\end{equation}
\begin{equation}
	\mathbf{M}^{\mathrm{e}}=\int_{A^{\mathrm{e}}}\rho\bm{N}^{\top}\mathrm{diag}(h,h^3/12,h^3/12)\bm{N}\mathrm{d}A
\end{equation}
\end{subequations}

By assembling all the element stiffness matrices and element mass matrices, we can obtain the global governing equation

\begin{equation}
	\mathbf{K}_0\mathbf{U}_0+\mathbf{M}_0\ddot{\mathbf{U}}_0=0
\end{equation}
in which $\mathbf{K}_0$ and $\mathbf{M}_0$ denote the global stiffness matrix and mass matrix, respectively.
Because of the periodic nature of phononic plate, the studying domain can be limited within a single unit cell on the basis of Floquet-Bloch theorem. Applying the Floquet-Bloch periodic condition in FEM analysis is equivalent to a nodal displacement transformation $\mathbf{U}=\mathbf{PU}_0$, where $\mathbf{P}$  is a $\bm{k}$-dependent matrix, as stated in our earlier work Ref.\cite{Wuying2018JSV}. The readers can also see \ref{sec:App.A} for more details. Under time-harmonic assumption, we have
\begin{align}
	\Big(\mathbf{K}(\bm{k})-\omega^2\mathbf{M}(\bm{k})\Big)\mathbf{U}=\mathbf{0}
	\label{eq:eigen_problem}
\end{align}
where $\mathbf{K}=\mathbf{P}^{\dag}\mathbf{K}_0\mathbf{P}, \mathbf{M}=\mathbf{P}^{\dag}\mathbf{M}_0\mathbf{P}$ are both Hermitian and $\bm{k}$-dependent matrices. Here $\dag$ represents Hermitian transpose. By solving the Hermitian eigenvalue problem equation (\ref{eq:eigen_problem}) with $\bm{k}$ varying along the edges of the first Irreducible Brillouin Zone (IBZ), we can obtain the dispersion relation $\omega=\omega(\bm{k})$.

\section{\label{sec:3}Topological Phase Transition}

\subsection{Band Inversion}

\begin{figure}[t]
	\centering
	\includegraphics[width=0.9\linewidth]{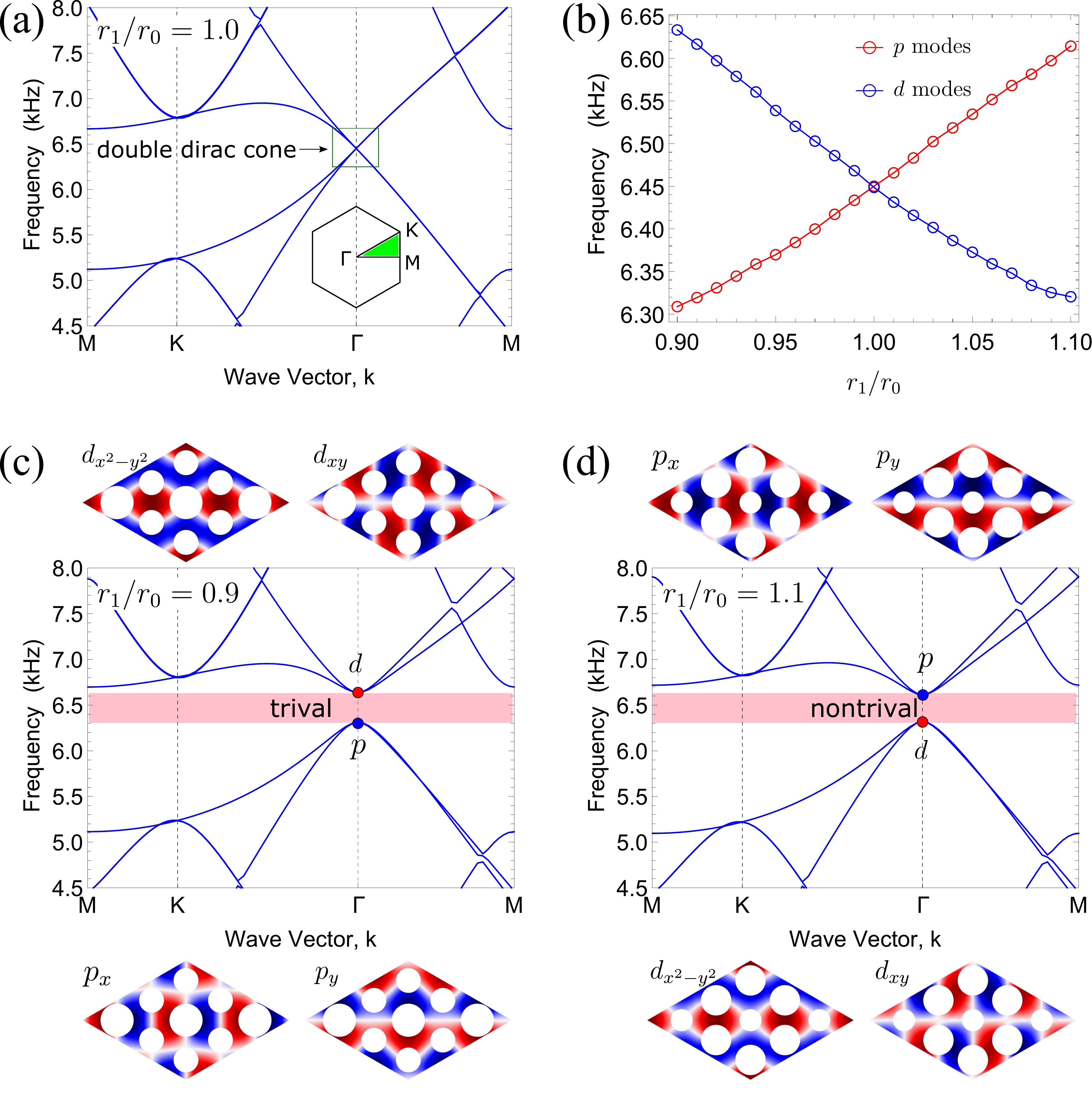}\\
	\caption{\textbf{Band structures of the phononic plates.} \textbf{(a)} A double Dirac cone emerges at the brillouin zone center when $r_1=r_0$. \textbf{(b)} The upper and lower bounds of the band gap vary versus $r_1$. \textbf{(c)} and \textbf{(d)} show the band structures and the eigenmodes when $r_1=0.9r_0$ and $r_1=1.1r_0$, respectively. When $r_1$ is perturbed, $r_2$ can be determined from the restriction condition $2r_1+r_2=3r_0$.}
	\label{fig:Fig2}
\end{figure}

Double Dirac cone plays an important role in imitating the quantum spin Hall effects in classical periodic systems. The emergence of a double Dirac cone at $\Gamma$ point can be achieved by varying the radii of perforated holes such that $r_1=r_2(=r_0)$ while their locations remain unchanged. When all the holes possess the same size, we observe a double Dirac cone at the BZ center. Figure \ref{fig:Fig2} (a) shows the band structure of the presented phononic plate in the case of $r_0=0.20a_0$. A four-fold degenerated state is observed at the frequency $f=6452\si{Hz}$ at $\Gamma$ point. Moreover, in the vicinity of $\Gamma$, four branches of dispersion curves touch at the degenerate point linearly (see \ref{A1}), in other words, a double Dirac cone has been formed. The construction of the double Dirac cone can be explained by zone-folding mechanism. When $r_1=r_2$, there are 9 uniform holes in the unit cell as shown in figure \ref{fig:Fig1}(b). In this case, the unit cell is actually a supercell, with $3\times 3$ primitive cells (the smallest unit cell). A (single) Dirac cone exists at the $K_0$ point (vertex of the BZ of the primitive cell) in the band structure of the primitive cell. By taking a $3\times 3$ supercell into consideration, the dispersion curves at $K_0$ point are folded into $\Gamma$ point twice. Therefore, a double Dirac cone is constructed at $\Gamma$ point in the BZ of the supercell. The four-fold degeneracy will be split into doubly degenerated dipolar states ($p$) and doubly degenerated quadripolar states ($d$) if the radii of the holes are tuned because of the brokenness of the translational symmetry. We adjust $r_1$ in the range of $[0.9r_0, 1.1r_0]$ with a restriction that $2r_1+r_2=3r_0$. The results plotted in figure \ref{fig:Fig2} (b) show that the gap between the $p$ and $d$ states closes and reopens as $r_1$ is tuned from $0.9r_0$ to $1.1r_0$. It is worth noting that $r_2=r_0$ when $r_1=r_0$ under the condition that $2r_1+r_2=3r_0$, therefore, the emergence of a double Dirac cone takes place.

As figure \ref{fig:Fig2} (b) shows, adjusting $r_1$ such that $r_1<r_0$ or $r_1>r_0$ can both open a band gap. Below we will show band inversion process by perturbing parameters $r_1$ from less to greater than $r_0$, which further leads to the topological transition from a topologically trival state to a non-trival state. Two cases are studied, case (I) for $r_1=0.9r_0$ and case (II) for $r_1=1.1r_0$, both breaking the double Dirac cone and opening a complete band gap, as shown in figure \ref{fig:Fig2} (c) and (d). Calculation results show that case (I) opens a gap ranged from $6314\sim 6639\si{Hz}$ and case (II) $6318\sim 6625\si{Hz}$. The band structures of both cases share almost the same profile and frequency ranges, but their corresponding eigenmodes are completely in contrast with each other. We plot the out of plane displacement distribution $w(x,y)$ of the eigenmodes corresponding to the $p$ and $d$ states in figure \ref{fig:Fig2} (c) and (d). For case (I), two modes below the gap behave like dipoles (marked by $p_x,~p_y$), and the modes above the gap behave like quadripolar (marked by $d_{x^2-y^2},~d_{xy}$). For case (II) the eigenmodes are flipped, that $p$ modes are above the gap and $d$ modes are below the gap. Here $p_x(p_y)$ represents the mirror symmetry of the eigenmodes along $x/y$ axis are even/odd(odd/even), and for $d_{x^2-y^2}(d_{xy})$ being  odd/odd (even/even). The band structure has experienced a process of gap closing to gap reopening, during which the eigenmodes above and below the gap at $\Gamma$ point have inversed.

\subsection{Topology of the Band Structures}
In this section we demonstrate that band inversion further induces the topological phase transition of the band structures. To reveal the topological property of the band gaps in figure \ref{fig:Fig2} (c) and (d), we employ the $\bm{k}\cdot\bm{p}$ perturbation method\cite{dresselhaus2007group} to construct an effective Hamiltonian for the proposed phononic plate, and further calculate the topological invariant, i.e., spin Chern number. The stiffness matrix $\mathbf{K}(\bm{k})$  in equation (\ref{eq:eigen_problem}) can be rewritten as the summation of $\mathbf{K}(\bm{k}_0)$ and a  perturbed term $\mathbf{K}(\Delta\bm{k})$ approximately by using the Taylor series. So equation (\ref{eq:eigen_problem}) can be expressed as
\begin{align}
	\big(\mathbf{K}(\bm{k}_0)+\mathbf{K}'-\omega^2\mathbf{M}(\bm{k}_0)\big)\mathbf{U}=\mathbf{0}
	\label{eq:S7}
\end{align}
in which $\mathbf{K}'\approx k_x\mathbf{K}_x+k_y\mathbf{K}_y$, and $\mathbf{K}_x, \mathbf{K}_y$ are constant matrices (that can be calculated from equation (\ref{eq:Reduce}). All the eigenvalues($\omega_{\bm{k}_0,n}^2$) and eigenvectors($\mathbf{U}_{\bm{k}_0,n}$) at $\Gamma$ point ($\bm{k}_0=\bm{0}$) can be calculated beforehand, so we can expand the eigenvector in equation (\ref{eq:eigen_problem}) into linear combination of $\mathbf{U}_{\bm{k}_0,n}$,
	\begin{equation}
		\mathbf{U} = \sum_{n}a_n\mathbf{U}_{\bm{k}_0,n}
		\label{eq:expansion}
	\end{equation}
where $a_n$ are expansion coefficients to be determined. Substituting equation (\ref{eq:expansion}) into equation (\ref{eq:eigen_problem}), and making use of the orthogonality relation $\mel{\mathbf{U}_{\bm{k}_0,m}}{\mathbf{M}}{\mathbf{U}_{\bm{k}_0,n}}=\delta_{mn}$, we can obtain that

	\begin{equation}
		(\mathcal{H}_0+\mathcal{H}'-\omega^2\mathbf{I})\{a\}=\mathbf{0}
		\label{eq:eigenprob}
	\end{equation}
in which $\mathcal{H}'_{mn}=\mel{\mathbf{U}_{\bm{k}_0,m}}{k_x\mathbf{K}_x+k_y\mathbf{K}_y}{\mathbf{U} _{\bm{k}_0,n}}$ is the first order perturbation term, and $\mathcal{H}_0=\text{diag}\{\omega_{\bm{k}_0,n}^2\}$ is a diagonal matrix. One should note that equation (\ref{eq:eigenprob}) still describes the eigenvalue problem in the entire space. Here we  develop an effective Hamiltonian to describe the dispersion around the two doubly degenerate states in a subspace, i.e., $\text{span}\{\ket{p_x},\ket{p_y},\ket{d_{x^2-y^2}},\ket{d_{xy}}\}$. Following the approach introduced in \cite{Wulonghua2015,Hecheng2016,Meijun2016}, we can obtain the matrix elements of the $4\times 4$ effective Hamiltonian on the basis $[p_x,p_y,d_{x^2-y^2},d_{xy}]$ as

\begin{equation}
\mathcal{H}_{mn}^{\text{eff}}=\mathcal{H}'_{mn}+\sum_{\alpha\neq m(n)}\frac{\mathcal{H}'_{m\alpha}\mathcal{H}'_{\alpha n}}{\omega^2_{\bm{k}_0,m(n)}-\omega^2_{\bm{k}_0,\alpha}}
\end{equation}

where the second term comes from the second order perturbation\cite{Wulonghua2015,Hecheng2016,Meijun2016}. With the FEM analysis, $\mathcal{H}_{mn}^{\mathrm{eff}}$ can be numerically calculated. Rewriting $\mathcal{H}^{\mathrm{eff}}$ on basis $[p_+,d_+,p_-,d_-]$ by an unitary transformation, we find that $\mathcal{H}^{\mathrm{eff}}$ can be expressed as
	\begin{align}
		\mathcal{H}^{\mathrm{eff}}=
		\begin{bmatrix}
			-M-Bk^2 & Ak_+ & 0 &0\\
			A^*k_- & M+Bk^2 &0 & 0\\
			0 & 0 & -M-B k^2 & Ak_-\\
			0 & 0 & A^*k_+ & M+B k^2
		\end{bmatrix}
		\label{eq:Heff}
	\end{align}
 with $p_{\pm}=(p_x\pm i p_y)/\sqrt{2},~ d_{\pm}=(d_{x^2-y^2}\pm i d_{xy})/\sqrt{2},~k_{\pm}=k_x\pm i k_y$, and $M=(\omega_{\bm{k}_0,d}^2-\omega_{\bm{k}_0,p}^2)/2$  is half of the gap-width between $p$ and $d$ modes. Coefficients $A$ and $B$ can be determined numerically following the above descriptions. The obtained effective Hamiltonian in equation (\ref{eq:Heff}) shares a similar form with Bernevig-Hughes-Zhang (BHZ) model for CdTe/HgTe/CdTe quantum well \cite{bernevig2006quantum}, indicating that our perforated phononic plate can behave a ``spin'' Hall effect. We can further calculate the spin Chern number\cite{Hecheng2016,Meijun2016}
	\begin{align}
		C=\pm\big(\mathrm{sgn}(M)+\mathrm{sgn}(B)\big)
	\end{align}

	Since the sign of $M$ changes from positive ($\omega_{\bm{k}_0,p}<\omega_{\bm{k}_0,d}$) to negative ($\omega_{\bm{k}_0,p}>\omega_{\bm{k}_0,d}$), and noticing the sign of $B$, which can be numerically evaluated, is typically negative, we can conclude that $MB<0$ before band inversion, thus $C_{\pm}=0$. While after the band inversion, we have $MB>0$ and $C_{\pm}=\pm 1$, indicating the topological transition from trival to non-trival band structure has taken place.

\section{\label{sec:4}Topological Edge States}

\begin{figure}[b]
	\centering
	\includegraphics[width=0.9\linewidth]{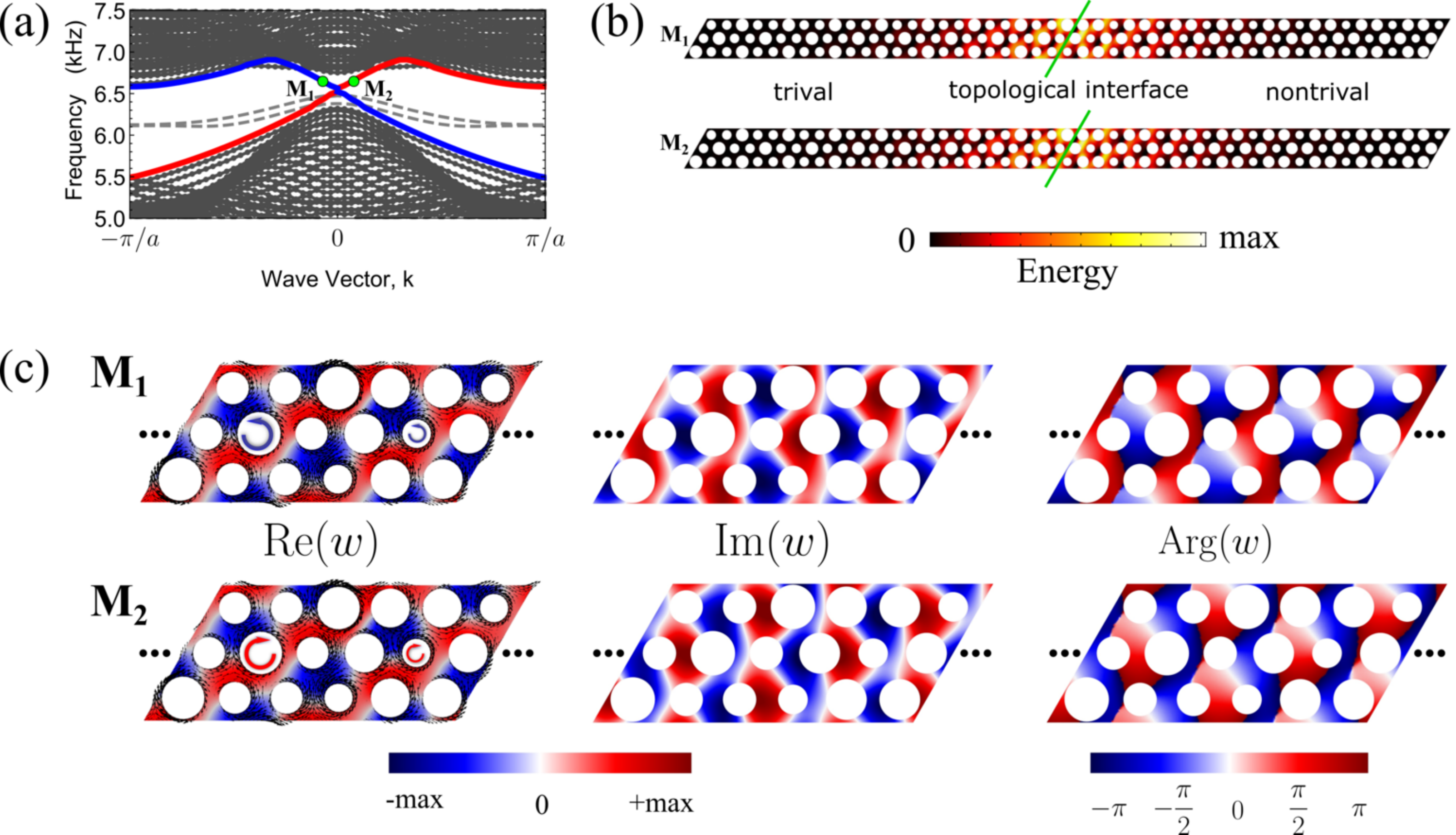}
	\caption{\textbf{(a)} Projected band structure of a ribbon-shaped supercell. The blue and red lines represent edge states, while the gray dot-solid lines represent bulk states. The green dots labelled as $M_1$ and $M_2$ are two specified edge states, with negative and positive group velocities, respectively. \textbf{(b)} shows the energy distributions for eigenmodes of states $M_1$ and $M_2$. It is evidently that the energy is localized at the interface between the topological trival and non-trival lattices. \textbf{(c)} shows the real parts, imaginary parts and the arguments for states $M_1$ and $M_2$ near the interface with magnified views. In the real-part plots, we also present the averaged time harmonic Poynting vectors, which represent the energy flows for both states. Energy flow for $M_1$ is anticlockwise, while energy flow for $M_2$ is clockwise, as shown by the thick blue and red circular arrows.}
	\label{fig:Fig3}
\end{figure}

To confirm the existence of topologically protected edge states for flexural waves in our system, we consider a ribbon-shaped supercell, constructed by 16 unit cells in $\bm{s}_2$ direction (8 topologically trival and 8 non-trival unit cells are arranged adjacently along the $\bm{s}_2$ direction, and in $\bm{s}_1$ direction  the system is still assumed to be periodic. Figure \ref{fig:Fig3} (a) shows the projected band structure of the supercell in $\Gamma K$ direction calculated from Mindlin's plate model. For pure lattice structure $\mathrm{I}$ (the topological trival lattice) or structure $\mathrm{II}$ (the topological non-trival lattice), there exists a complete band gap and the two gaps share a common frequency range $[6318\si{Hz},6625\si{Hz}]$, but a pair of topological edge states (blue and red solid lines) exist evidently in the overlapped bulk band gap frequency regime of the two types of lattices. It is also worth of noting that there exist two boundary modes in the band gap range, as shown in dashed gray lines in Figure \ref{fig:Fig3} (a). We take two specified edge states as an example. As figure \ref{fig:Fig3} (a) shows, the two states with opposite group velocities belong to different branches  of these edge states (marked by state $M_1$ and state $M_2$). We plot the square of the amplitude $|w(x,y)|^2$ (representing the vibration energy) in figure \ref{fig:Fig3} (b), which unambiguously demonstrates that $M_1$ and $M_2$ states are localized at the lattice-interface since the deformation amplitude decreases rapidly with the distance away from the lattice interface. Figure \ref{fig:Fig3} (c) supports a magnified view for two adjacent unit cells at the interface. We plot the real parts $\mathrm{Re}(w)$, imaginary parts $\mathrm{Im}(w)$ and the arguments $\mathrm{Arg}(w)$ for states $M_1$ and $M_2$, and find that $\mathrm{Re}(w_{M_1})=\mathrm{Re}(w_{M_2}),~\mathrm{Im}(w_{M_1})=-\mathrm{Im}(w_{M_2})$ and $\mathrm{Arg}(w_{M_1})=-\mathrm{Arg}(w_{M_2})$. This finding, especially the argument information, gives us guidance
about the selectively excitation of a one-way edge mode with multiple sources, as we will discuss in the next section. We also plot the time-averaged Poynting vectors of two adjacent unit cell at the interface by black arrows. The power flow of the center regions of each unit cell is anti-clockwise for state $M_1$, and clockwise for state $M_2$, unveiling the pseudo spin up and down characteristics of each state.
When we introduce complex number in elastic wave propagation problems, physical quantities like displacements, velocities and accelerations, should take only the real (or imaginary) part. Since we observed $\mathrm{Re}(w_{M_1})=\mathrm{Re}(w_{M_2}),~\mathrm{Im}(w_{M_1})=-\mathrm{Im}(w_{M_2})$, and do not forget eigen-vectors allow a constant scale ratio difference, so whichever part (real or imaginary) we take, we conclude that the eigenvector of state $M_1$ and $M_2$ are the ``same''. But for physical quantities like energy and power, we have to take both the real and imaginary parts into consideration\cite{achenbach_wave}. So states $M_1$ and $M_2$ possessing the same real part and opposite imaginary part is the reason why the Poynting vectors (energy flow) for these two states are pointing to the opposite directions.

\section{\label{sec:5}Robust One-way Edge Wave Propagation}

\subsection{\label{sec:oneway}Topologically Directional Wave Guiding}

\begin{figure}[t]
	\centering
	\includegraphics[width=0.8\linewidth]{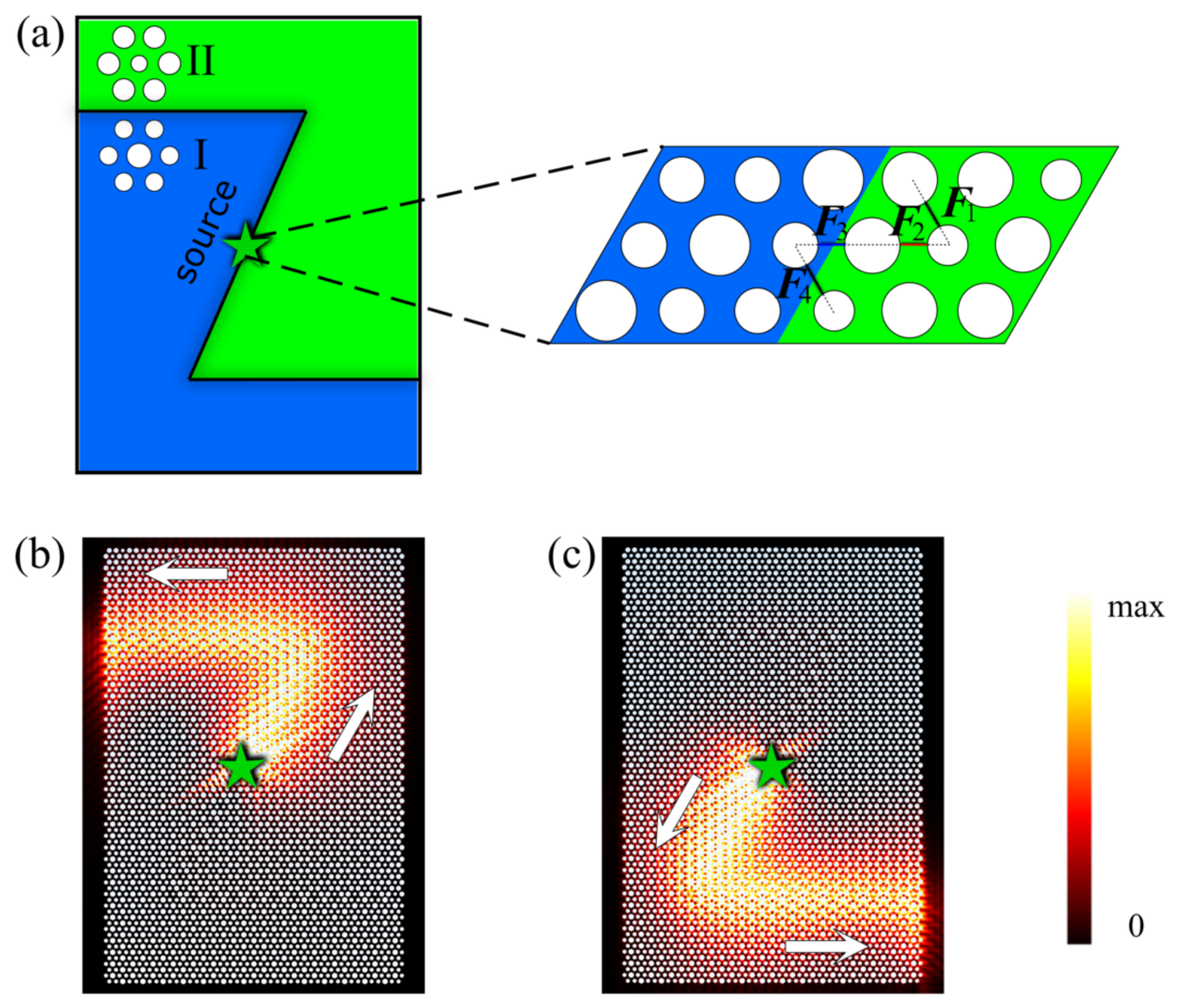}
	\caption{\textbf{Unidirectional propagation of topologically protected edge states.} \textbf{(a)} Schematic view of the present  wave guide. Two types of lattices are separated by a ``Z'' shape domain wall. Multi-line-excitations are utilized to selectively excite a particular edge state, with $F_1=F_4=F_0,F_2=F_0\exp(\pm i\pi/2),F_3=F_0\exp(\mp i\pi/2)$. \textbf{(b)} up going edge state (positive group velocity) and \textbf{(c)} down going edge state (negative group velocity) are selectively excited, by taking the upper/lower sign in the expressions of $F_2$ and $F_3$.}
	\label{fig:Fig4}
\end{figure}

Full wave simulations for flexural waves propagation in a finite lattice ($15\times 26$ cells) is carried out via COMSOL Multiphysics, a finite element solver. as shown in figure \ref{fig:Fig4}. We use the Solid Mechanics Module to model our plate structure, and adopt the Frequency Domain Analysis to simulate the time-harmonic flexural wave propagation. The present sample is composed of two different lattices (type I and type II), separated by a ``Z'' shape domain wall. The excitation sources are arranged inside the lattice sample, which is surrounded by perfectly matched layers (PML) so that the output waves cannot be reflected back into the sample. Edge states show up in pairs in the bulk band gap frequency range, which increases the difficulty to excite an unidirectional transport. We utilize multiple sources with phase differences to overcome this challenge, i.e., to selectively excite the directional up (down) going topological edge state. The harmonic force sources $F_i=F_0\exp(i\varphi_i)$ are shown in the right panel of figure \ref{fig:Fig4} (a). Note that the time harmonic term $\exp(-\omega t)$ is omitted here, and all the forces possess the same amplitude $F_0$ but different phases,
\begin{equation}
	\varphi_1=\varphi_4=0,\quad\varphi_2=\pm\pi/2,\quad\varphi_3=\mp\pi/2
    \label{eq:phase}
\end{equation}
The phases of the excitation sources are determined by extracting the argument information of the edge states (see figure \ref{fig:Fig3} (c)). When taking the upper sign in equation (\ref{eq:phase}), the excitation phases are $\varphi_1=\varphi_4=0,\varphi_2=\pi/2,\varphi_3=-\pi/2$, which is consistent with the argument of state $M_2$, as we can see in figure \ref{fig:Fig3} (c). The simulated wave field With excitation frequency $f=6500\si{Hz}$, which is exactly the frequency for state $M_2$, is plotted in figure \ref{fig:Fig4} (b). The wave travels only in a single direction with positive group velocity, being consistent with state $M_2$. In this case we successfully excited only \textit{one-way} edge mode. Similarly, when taking the lower sign in equation (\ref{eq:phase}), we would expect another \textit{one-way} edge mode with negative group velocity be excited at frequency of $6500\si{Hz}$. In other words, edge state $M_1$ is selective excited, which is confirmed by the wave propagation simulation shown in \ref{fig:Fig4} (c).

\subsection{Robust Wave Propagation}

\begin{figure}[t]
	\centering
	\includegraphics[width=0.8\linewidth]{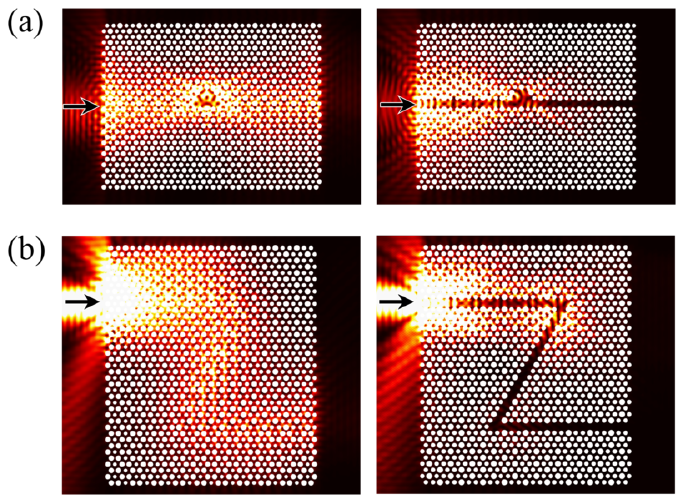}
	\caption{\textbf{Wave field of topological (left) and conventional (right) wave guides with different  defects.} Excitation frequency is chosen to be $6500\si{Hz}$.
	%Point defects near the middle part of the samples (a) and Sharp bends (b) are introduced.
	}
	\label{fig:Fig5}
\end{figure}

The $C_{6v}$ symmetry protected flexural wave propagation of our topological wave guide is studied by intentionally introducing different defects. For comparison, wave propagation in conventional wave guides with the same defects is also studied. Topological wave guide studied here is constructed by the topological trival lattice (structure I) and non-trival lattice (structure II), as addressed in section \ref{sec:oneway}. Conventional wave guide is realized by simply removing a set of holes along a desired path in pure lattice I.
The incident waves are generated by a line-excitation from the left side of the sample, and the force amplitude along the excitation line is set to be Gaussian distribution, which can be expressed as $F=F_0\exp(-\frac{(y-y_0)^2}{4b^2})$. Here $y_0$ determines the wave-beam center, and parameter $b$ determines the width of the wave-beam.
The first example is a straight wave guide with additional point defects, as figure \ref{fig:Fig5} (a) shows. For topological wave guide, the normally incident waves can travel through the entire sample, and no backscattering is observed obviously at the point defect. As for the conventional wave guide, strong backscattering are observed so that no output waves can be detected at the right port. The second example studies the wave propagating along a path with sharp bends for both types of structures. Simulation results in figure \ref{fig:Fig5} (b) show that waves in topological lattice are backscattering free against sharp bends while waves in conventional wave guides are not.

\section{\label{sec:6}Conclusions}
In conclusion, we have proposed a perforated phononic plate and studied the pseudo-spin states for flexural waves propagation based on Mindlin's plate theory. A double Dirac cone is observed at $\Gamma$ by carefully adjusting the size of perforated holes such that all the holes are uniform, which can be explained by the zone-folding mechanism. When the radii are perturbed in different directions, the double Dirac cone opens because the translational symmetry is broken. We have shown that a topological phase transition takes place due to the band inversion by perturbing the radii of the holes while preserving the $C_{6v}$ point group symmetry. An effective Hamiltonian is developed to describe the topological properties of the band structures and to calculate topological invariants, i.e., the spin Chern number. Zero/non-zero Chern numbers are obtained before/after the band inversion, which again confirms that the topological transition has taken place.  In addition, the projected dispersion of a ribbon-shaped finite supercell with topologically trival and non-trival cells arranged adjacently is investigated. Result demonstrates the existence of a pair of topological edge states located within the bulk gap range, which support two pseudo-spin states. With full wave simulations, we demonstrate the robust unidirectional transport of topologically protected edge states for flexural waves. Observation in this paper paves a new way for the studies and practical applications in wave guiding, vibration isolating,  wave filtering and related fields. Further studies including the experimental verification of edge wave propagation in the proposed metamaterial plate will be reported in authors' future publications.

\section*{Acknowledgements}
We acknowledge Rajesh Chaunsali (University of Washington) gratefully for fruitful discussions.

\appendix

\section{\label{sec:App.A}Applying Bloch Condition in FEM}
% Displacement components for flexural wave modes in a flat plate are expressed by
% \begin{equation}
% \left\{
% 	\begin{aligned}
% 		u(x,y,z)&=z\psi_x(x,y)\\
%         v(x,y,z)&=z\psi_y(x,y)\\
%         w(x,y,z)&=w(x,y)
% 	\end{aligned}
% \right.
% \label{eq:Asump}
% \end{equation}
% where $u,v,w$ are the displacements of the plate along $x,y$ and $z$ axis, and $\psi_{x(y)}$ is the rotation angle of the norm vector with respect to $x(y)$ axis. Besides the $O-xyz$ coordinate system notation, we employ another notation system $O-x_1x_2x_3$ in which direction $x_1,x_2,x_3$ represent $x,y,z$. So the displacement components $u,v,w$ are mapped to $u_1,u_2,u_3$ in a similar way. With this notation, the strain tensor can be written as
% \begin{equation}
% 	\varepsilon_{ij}=\frac{\partial u_i}{\partial x_j}+\frac{\partial u_j}{\partial x_i}
% \end{equation}

% It is worth noting that $\varepsilon_{33}(\varepsilon_z)$ valishes under the assumption of the Mindlin's plate theory equation (\ref{eq:Asump}). The stress-strain relations are
% \begin{equation}
% 	\sigma_b=D_b\varepsilon_b,\quad \sigma_s=D_s\varepsilon_s
% \end{equation}

For the dispersion calculation of periodic systems, we should apply the Bloch-Floquet conditions on the periodic edge pairs of the unit cell. As figure A1 shows, all the finite element nodes are grouped into 9 node sets. Node set 1 includes all the nodes locate within the inner domain, and node sets 2 -- 5 the nodes on the four edges (but not including the endpoints of the edges). The rest 4 nodes at the unit cell corners are node sets 6 -- 9. We would like to note that the current mesh shown in Fig. 1A  (a) may be a little coarse that the results may not meet the convergence condition, thus a finer mesh as Fig. 1A (b) shows is required to check the convergence of FE analysis.

\begin{figure}[b]
	\centering
	\includegraphics[width=0.9\linewidth]{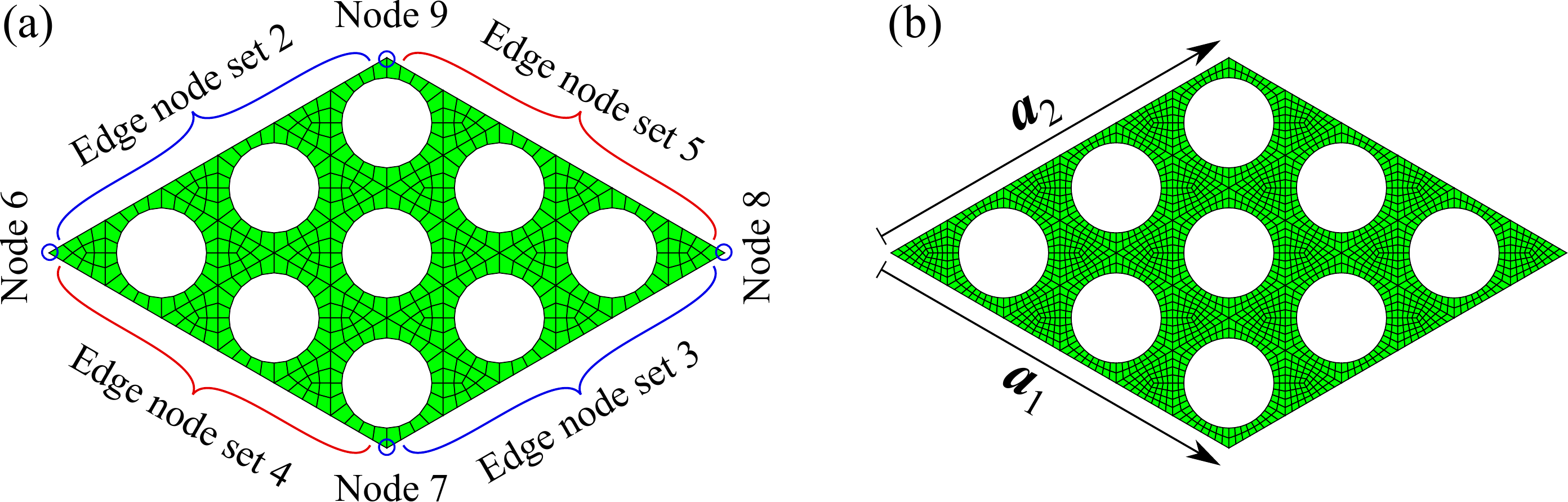}
	\caption{FE meshes of the unit cell. (a) A coarse mesh to better illustrate the 9 node sets: node sets 2 -- 9 are nodes on the boundaries and node set 1 are those locate in the inner domain. (b) A finer mesh is utilized in the FE analysis to ensure the convergence of results.}
	\label{fig:Fig6}
\end{figure}

The Bloch-Floquet conditions states that
\begin{equation}
\begin{aligned}
	\mathbf{u}_3&=\exp(i\bm{k}\cdot\bm{a}_1)\mathbf{u}_2,\quad \mathbf{u}_5=\exp(i\bm{k}\cdot\bm{a}_2)\mathbf{u}_4\\
	u_7&=\exp(i\bm{k}\cdot\bm{a}_1)u_6,\quad u_9=\exp(i\bm{k}\cdot\bm{a}_2)u_6\\
	u_8&=\exp(i\bm{k}\cdot(\bm{a}_1+\bm{a}_2))u_6
\end{aligned}
\label{eq:Reduce}
\end{equation}
in which $\bm{k}$ is wave vector, $\bm{a}_1,\bm{a}_2$ are the two lattice vectors. From equation (\ref{eq:Reduce}) we can reduce the displacement vector $\mathbf{U}_0$ which consists of all of the nodal displacements to a ``reduced'' vector $\mathbf{U}$ whose components are the nodal displacements of node sets 1, 2, 4 and 6, expressed as
\begin{align}
	\mathbf{U}_0=\mathbf{P}\mathbf{U}
	\label{eq:trans}
\end{align}
where $\mathbf{U}_0=[\mathbf{u}_1,\mathbf{u}_2,\mathbf{u}_3,\mathbf{u}_4,\mathbf{u}_5,u_6,u_7,u_8,u_9]^{\top}$, $\mathbf{U}=[\mathbf{u}_1,\mathbf{u}_2,\mathbf{u}_4,u_6]^{\top}$, and matrix $\mathbf{P}$ is called the transformation matrix, whose components can be obtained from equation (\ref{eq:Reduce}).
For the dispersion analysis of periodic structures, by applying Bloch-Floquet conditions equation (\ref{eq:Reduce}) or equation (\ref{eq:trans}), one can obtain the eigenvalue problem equation (\ref{eq:eigen_problem}).

\section{Comparison Between Plate Model and Solid Model}

\begin{figure}[hbtp]
	\centering
	\includegraphics[width=\linewidth]{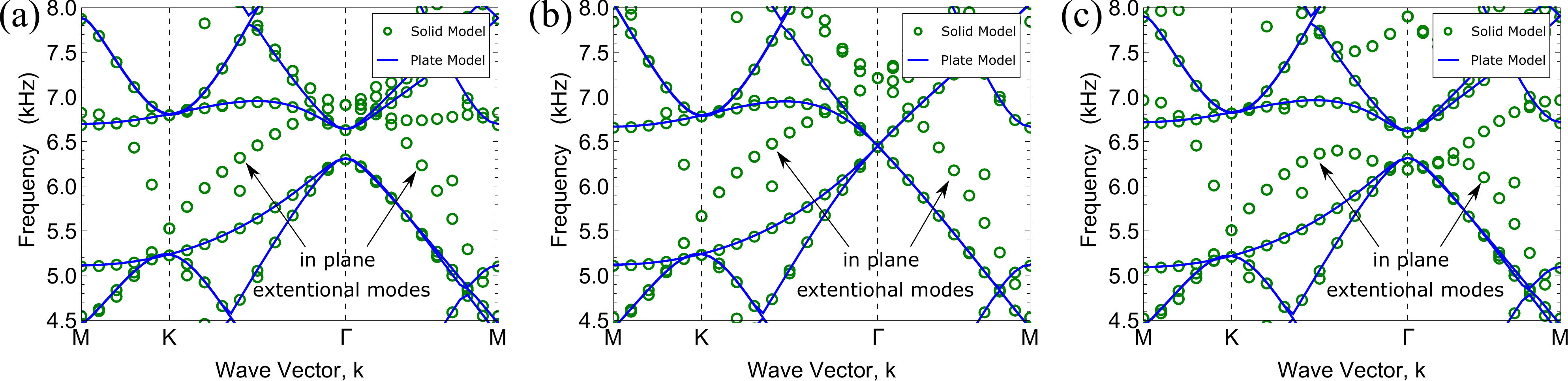}
	\caption{\textbf{Validity  of our FEM for flexural wave modes analysis}. Dispersion curves  obtained from Mindlin's plate theory (blue lines) match well with those obtained from elastodynamics equation (green circles). (a)-(c) represent the band structures for $r_1/r_0=0.9,1.0$ and $1.1$.}
	\label{fig:Fig7}
\end{figure}

In this section, we will check the accuracy of the proposed FEM for flexural wave analysis. Both 2D model based on Mindlin's plate theory and 3D model based on elastic dynamic equation
\begin{equation}
(\lambda+\mu)\nabla\nabla\cdot\mathbf{u}+\mu\nabla^2\mathbf{u}=\rho\ddot{\mathbf{u}}
\end{equation}
are utilized to analyse the dispersion of the present plate. The blue solid lines in figure \ref{fig:Fig7} represent the dispersion curves obtained from Mindlin's plate theory model, and the green circles represent dispersion curves of elastic solid model. By comparing these two set of results we can find that the plate model can accurately describe the flexural wave modes for a thick plate and efficiently eliminate other types of wave modes (SH modes and in-plane extensional modes) which are not our research interests in this paper.

\section{\label{A1}Effective Hamiltonian near The Double Dirac Cone}

We employ the degenerate second-order $\bm{k}\cdot\bm{p}$ perturbation theory\cite{dresselhaus2007group} to calculate the effective Hamiltonian around $\Gamma$ point. For a fixed plate structure, for example, $r_1=r_2$ or $r_1>r_2$ or $r_1<r_2$, the eigenvalues $\omega^2_{\bm{k}_0,n}$ and eigenvectors $U_{\bm{k}_0,n}$ at $\bm{k}_0=0$ are calculated beforehand. Small wave vector $\bm{k}$ (around $\bm{k}_0$) is treated as a perturbation. Our goal is to find the effective Hamiltonian when wave vector takes $\bm{k}$, with only $\omega^2_{\bm{k}_0,n}$ and $U_{\bm{k}_0,n}$ are already known. According to the degenerate second-order $\bm{k}\cdot\bm{p}$ perturbation theory, the matrix components of the effective Hamiltonian at $\bm{k}$ point is given by (see page 316--319 in Ref.\cite{dresselhaus2007group} for more details about the derivation)
\begin{equation}
\mathcal{H}_{mn}^{\text{eff}}=\mathcal{H}'_{mn}+\sum_{\alpha}^{\infty}\frac{\mathcal{H}'_{m\alpha}\mathcal{H}'_{\alpha n}}{\omega^2_{\bm{k}_0,m(n)}-\omega^2_{\bm{k}_0,\alpha}}
\end{equation}

in which subscripts $m,n$ denote the degenerate states that we are interested in, and subscript $\alpha$ denotes all of the rest states. We recall that $\mathcal{H}'_{mn}=\mel{\mathbf{U}_{\bm{k}_0,m}}{k_x\mathbf{K}_x+k_y\mathbf{K}_y}{\mathbf{U} _{\bm{k}_0,n}}$ is the first order perturbation term and $K_x$ and $K_y$ are constant matrices.

In this section, we consider the case of $r_1=r_2$. Since a double Dirac cone (four-fold degenerate states) emerges, $m,n$ should be taken over the corresponding 4 states indexes and $\alpha$ the rest. We should also note that the other states ($\alpha$) are far away from the double Dirac cone, so $|\omega^2_{\bm{k}_0,m(n)}-\omega^2_{\bm{k}_0,\alpha}|$ is very large that the summation term over $\alpha$ can be neglected (only when $r_1=r_2$, but not for the case $r_1\neq r_2$).  So we have

\begin{figure}[b]
	\centering
	\includegraphics[width=0.5\linewidth]{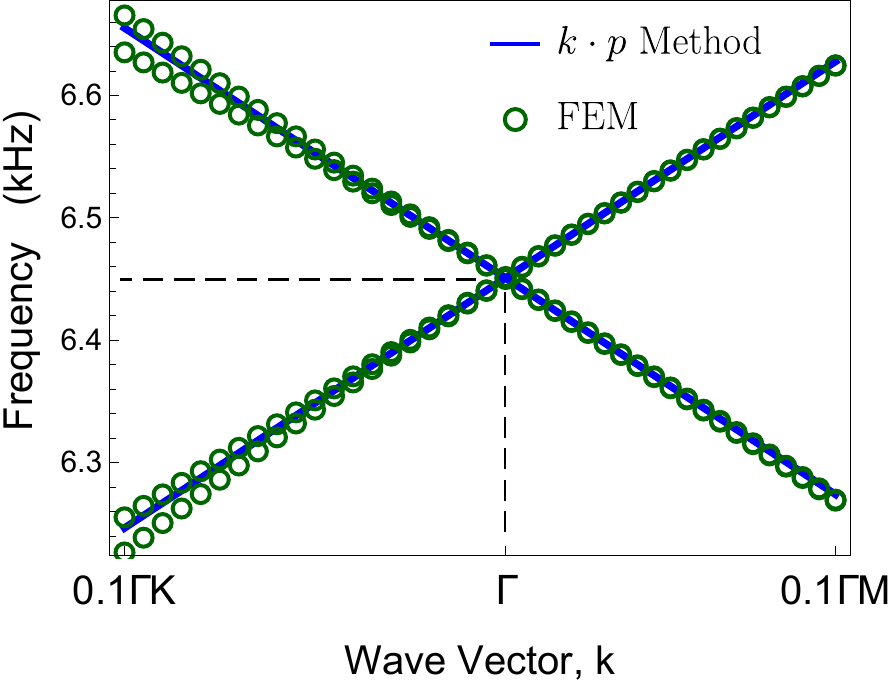}
	\caption{Double Dirac Cone near $\Gamma$ ($r_1=r_2=0.20a$). Solid lines are predicted from the effective Hamiltonian and the green circles are calculated by FEM.}
	\label{fig:Fig8}
\end{figure}
\begin{equation}
\mathcal{H}_{mn}^{\text{eff}}=\mathcal{H}'_{mn}=\mel{\mathbf{U}_{\bm{k}_0,m}}{k_x\mathbf{K}_x+k_y\mathbf{K}_y}{\mathbf{U} _{\bm{k}_0,n}}
\label{eq:CompnEH}
\end{equation}

Eq.\eqref{eq:CompnEH} gives the expressions of each matrix components of $\mathcal{H}^{\rm{eff}}$. By numerical calculation, we find that $\mathcal{H}^{\text{eff}}$ can be expressed as

\begin{align}
\mathcal{H}^{\text{eff}}=1.0\mathrm{e}7
	\begin{bmatrix}
		0 & -A & -B & C\\
		A & 0  & -C & -B\\
		B & C & 0 & A\\
		-C & B & -A & 0
	\end{bmatrix}
	\label{eq:Heff_DoubleDirac}
\end{align}
where $A=(0.3075 k_x-0.2556 k_y)i,B=(0.1016 k_x+2.1398 k_y)i,C=(2.13 k_x-0.0637 k_y)i$.

Eigenvalues of equation (\ref{eq:Heff_DoubleDirac}) can be solved explicitly,
\begin{eqnarray}
	\varepsilon_{1,2}&=-i\sqrt{A^2+B^2+C^2}&\approx +2.15\mathrm{e}7k\nonumber\\
	\varepsilon_{3,4}&=+i\sqrt{A^2+B^2+C^2}&\approx -2.15\mathrm{e}7k
	 \label{eq:S11}
\end{eqnarray}

in which $\varepsilon=\omega^2-\omega_0^2$. so equation (\ref{eq:S11}) can be rewritten as
\begin{align}
	\omega^2-\omega_0^2\approx 2\omega_0\Delta\omega=\pm 2.15\mathrm{e}7(k-0)
\end{align}

so the slopes of the dispersion curves near the double Dirac cone are given by
\begin{align}
	\frac{\Delta\omega}{\Delta k}=\pm\frac{2.15\mathrm{e}7}{2\omega_0}
	\label{eq:slope}
\end{align}

We note that each root is two-fold, indicating the two cones are uniform. Around $\Gamma$ point, dispersions obtained from our $\bm{k}\cdot \bm{p}$ method show excellent agreement with those from FEM, as shown in figure {\ref{fig:Fig8}}.

%\begin{figure*}[!t]
%	\centering
%	\includegraphics[width=0.9\linewidth]{figures/FigS6.pdf}
%	\caption{$p_x,p_y$ and $d_{x^2-y^2},d_{xy}$ modes constructed by linear combinations of $p_1,p_2$ and $d_1,d_2$ modes.}
%	\label{fig:S6}
%\end{figure*}

\section*{References}
\bibliography{References}% Produces the bibliography via BibTeX.

\begin{thebibliography}{10}

\bibitem{von1986quantized}
Klaus von Klitzing.
\newblock The quantized hall effect.
\newblock {\em Rev. Mod. Phys.}, 58:519--531, Jul 1986.

\bibitem{hasan2010colloquium}
M.~Z. Hasan and C.~L. Kane.
\newblock Colloquium.
\newblock {\em Rev. Mod. Phys.}, 82:3045--3067, Nov 2010.

\bibitem{qi2011topological}
Xiao-Liang Qi and Shou-Cheng Zhang.
\newblock Topological insulators and superconductors.
\newblock {\em Rev. Mod. Phys.}, 83:1057--1110, Oct 2011.

\bibitem{haldane2008}
F.~D.~M. Haldane and S.~Raghu.
\newblock Possible realization of directional optical waveguides in photonic
  crystals with broken time-reversal symmetry.
\newblock {\em Phys. Rev. Lett.}, 100:013904, Jan 2008.

\bibitem{lu2014}
Ling Lu, John~D Joannopoulos, and Marin Solja{\v{c}}i{\'c}.
\newblock Topological photonics.
\newblock {\em Nat. Photon.}, 8(11):821--829, 2014.

\bibitem{Wulonghua2015}
Long-Hua Wu and Xiao Hu.
\newblock Scheme for achieving a topological photonic crystal by using
  dielectric material.
\newblock {\em Phys. Rev. Lett.}, 114:223901, Jun 2015.

\bibitem{khanikaev2015}
Alexander~B Khanikaev, Romain Fleury, S~Hossein Mousavi, and Andrea Al{\`u}.
\newblock Topologically robust sound propagation in an angular-momentum-biased
  graphene-like resonator lattice.
\newblock {\em Nat. Commun.}, 6, 2015.

\bibitem{Yangzhaoju2015}
Zhaoju Yang, Fei Gao, Xihang Shi, Xiao Lin, Zhen Gao, Yidong Chong, and Baile
  Zhang.
\newblock Topological acoustics.
\newblock {\em Phys. Rev. Lett.}, 114:114301, Mar 2015.

\bibitem{Chenzeguo2016}
Ze-Guo Chen and Ying Wu.
\newblock Tunable topological phononic crystals.
\newblock {\em Phys. Rev. Applied}, 5:054021, May 2016.

\bibitem{Hecheng2016}
Cheng He, Xu~Ni, Hao Ge, Xiao-Chen Sun, Yan-Bin Chen, Ming-Hui Lu, Xiao-Ping
  Liu, and Yan-Feng Chen.
\newblock Acoustic topological insulator and robust one-way sound transport.
\newblock {\em Nat. Phys.}, 12(12):1124--1129, 2016.

\bibitem{Meijun2016}
Jun Mei, Zeguo Chen, and Ying Wu.
\newblock Pseudo-time-reversal symmetry and topological edge states in
  two-dimensional acoustic crystals.
\newblock {\em Sci. Rep.}, 6, 2016.

\bibitem{Xiabaizhan2017}
Bai-Zhan Xia, Ting-Ting Liu, Guo-Liang Huang, Hong-Qing Dai, Jun-Rui Jiao,
  Xian-Guo Zang, De-Jie Yu, Sheng-Jie Zheng, and Jian Liu.
\newblock Topological phononic insulator with robust pseudospin-dependent
  transport.
\newblock {\em Phys. Rev. B}, 96:094106, Sep 2017.

\bibitem{SimonYves2017}
Simon Yves, Romain Fleury, Fabrice Lemoult, Mathias Fink, and Geoffroy Lerosey.
\newblock Topological acoustic polaritons: robust sound manipulation at the
  subwavelength scale.
\newblock {\em New J. Phys.}, 19(7):075003, 2017.

\bibitem{Lujiuyang2016}
Jiuyang Lu, Chunyin Qiu, Manzhu Ke, and Zhengyou Liu.
\newblock Valley vortex states in sonic crystals.
\newblock {\em Phys. Rev. Lett.}, 116:093901, Feb 2016.

\bibitem{fleury2016}
Romain Fleury, Alexander~B Khanikaev, and Andrea Al{\`u}.
\newblock Floquet topological insulators for sound.
\newblock {\em Nat. Commun.}, 7, 2016.

\bibitem{Lujiuyang2017}
Jiuyang Lu, Chunyin Qiu, Liping Ye, Xiying Fan, Manzhu Ke, Fan Zhang, and
  Zhengyou Liu.
\newblock Observation of topological valley transport of sound in sonic
  crystals.
\newblock {\em Nat. Phys.}, 13(4):369--374, 2017.

\bibitem{Zhangzhiwang2017}
Zhiwang Zhang, Qi~Wei, Ying Cheng, Ting Zhang, Dajian Wu, and Xiaojun Liu.
\newblock Topological creation of acoustic pseudospin multipoles in a flow-free
  symmetry-broken metamaterial lattice.
\newblock {\em Phys. Rev. Lett.}, 118:084303, Feb 2017.

\bibitem{NiXiang2017}
Xiang Ni, Maxim~A Gorlach, Andrea Alu, and Alexander~B Khanikaev.
\newblock Topological edge states in acoustic kagome lattices.
\newblock {\em New J. Phys.}, 19(5):055002, 2017.

\bibitem{susstrunk2015}
Roman S{\"u}sstrunk and Sebastian~D. Huber.
\newblock Observation of phononic helical edge states in a mechanical
  topological insulator.
\newblock {\em Science}, 349(6243):47--50, 2015.

\bibitem{Wangpai2015}
Pai Wang, Ling Lu, and Katia Bertoldi.
\newblock Topological phononic crystals with one-way elastic edge waves.
\newblock {\em Phys. Rev. Lett.}, 115:104302, Sep 2015.

\bibitem{nash2015}
Lisa~M. Nash, Dustin Kleckner, Alismari Read, Vincenzo Vitelli, Ari~M. Turner,
  and William T.~M. Irvine.
\newblock Topological mechanics of gyroscopic metamaterials.
\newblock {\em Proc. Natl. Acad. Sci. U.S.A.}, 112(47):14495--14500, 2015.

\bibitem{huber2016}
Sebastian~D Huber.
\newblock Topological mechanics.
\newblock {\em Nat. Phys.}, 12(7):621--623, 2016.

\bibitem{torrent2013}
Daniel Torrent, Didier Mayou, and Jos\'e S\'anchez-Dehesa.
\newblock Elastic analog of graphene: Dirac cones and edge states for flexural
  waves in thin plates.
\newblock {\em Phys. Rev. B}, 87:115143, Mar 2013.

\bibitem{mousavi2015}
S~Hossein Mousavi, Alexander~B Khanikaev, and Zheng Wang.
\newblock Topologically protected elastic waves in phononic metamaterials.
\newblock {\em Nat. Commun.}, 6, 2015.

\bibitem{miniaci2017}
M.~{Miniaci}, R.~K. {Pal}, B.~{Morvan}, and M.~{Ruzzene}.
\newblock {Observation of topologically protected helical edge modes in Kagome
  elastic plates}.
\newblock {\em ArXiv e-prints}, October 2017.

\bibitem{Yusiyuan2017}
S.-Y. {Yu}, C.~{He}, Z.~{Wang}, F.-K. {Liu}, X.-C. {Sun}, Z.~{Li}, M.-H. {Lu},
  X.-P. {Liu}, and Y.-F. {Chen}.
\newblock {A Monolithic Topologically Protected Phononic Circuit}.
\newblock {\em ArXiv e-prints}, July 2017.

\bibitem{chaunsali2018}
Rajesh Chaunsali, Chun-Wei Chen, and Jinkyu Yang.
\newblock Subwavelength and directional control of flexural waves in
  zone-folding induced topological plates.
\newblock {\em Phys. Rev. B}, 97:054307, Feb 2018.

\bibitem{foehr2017}
A.~{Foehr}, O.~R. {Bilal}, S.~D. {Huber}, and C.~{Daraio}.
\newblock {Spiral-based phononic plates: From wave beaming to topological
  insulators}.
\newblock {\em ArXiv e-prints}, December 2017.

\bibitem{pal2017edge}
Raj~Kumar Pal and Massimo Ruzzene.
\newblock Edge waves in plates with resonators: an elastic analogue of the
  quantum valley hall effect.
\newblock {\em New J. Phys.}, 19(2):025001, 2017.

\bibitem{Zhuhongfei2017}
H.~{Zhu}, T.-W. {Liu}, and F.~{Semperlotti}.
\newblock {Design and Experimental Observation of Valley-Hall Edge States in
  Diatomic-Graphene-like Elastic Waveguides}.
\newblock {\em ArXiv e-prints}, December 2017.

\bibitem{Wuying2018}
Ying Wu, Rajesh Chaunsali, Hiromi Yasuda, Kaiping Yu, and Jinkyu Yang.
\newblock Dial-in topological metamaterials based on bistable stewart platform.
\newblock {\em Sci. Rep.}, 8(1):112, 2018.

\bibitem{achenbach_wave}
Jan Achenbach.
\newblock {\em Wave propagation in elastic solids}, volume~16.
\newblock Elsevier, 2012.

\bibitem{ferreira_matlab}
{\em Analysis of Mindlin plates}, pages 161--201.
\newblock Springer Netherlands, Dordrecht, 2009.

\bibitem{Wuying2018JSV}
Ying Wu, Kaiping Yu, Linyun Yang, Rui Zhao, Xiaotian Shi, and Kuo Tian.
\newblock Effect of thermal stresses on frequency band structures of elastic
  metamaterial plates.
\newblock {\em J. Sound Vib.}, 413:101 -- 119, 2018.

\bibitem{dresselhaus2007group}
Mildred~S Dresselhaus, Gene Dresselhaus, and Ado Jorio.
\newblock {\em Group theory: application to the physics of condensed matter}.
\newblock Springer Science \& Business Media, 2007.

\bibitem{bernevig2006quantum}
B~Andrei Bernevig, Taylor~L Hughes, and Shou-Cheng Zhang.
\newblock Quantum spin hall effect and topological phase transition in hgte
  quantum wells.
\newblock {\em Science}, 314(5806):1757--1761, 2006.

\end{thebibliography}

\end{document}